\documentclass[9pt,conference]{IEEEtran}
\usepackage[]{waspaa25}
\usepackage{arydshln}

\usepackage{waspaa25} %

\usepackage{bm} %
\usepackage{color}
\usepackage{pifont}%
\usepackage{caption}
\usepackage{float}
\usepackage{amsmath}
\usepackage{multirow, multicol}
\newcommand{\cmt}[1]{\ignorespaces}

\title{Balancing \cmt{the Trade-offs Between} Information Preservation and Disentanglement in Self-Supervised Music Representation Learning}

\name{Julia Wilkins$^{1}$,
      Sivan Ding$^{1}$,
      Magdalena Fuentes$^{1}$,
      Juan Pablo Bello$^{1}$}
\address{$^{1}$New York University, New York, USA \;
}

\begin{document}

\maketitle

\begin{abstract}
Recent advances in self-supervised learning (SSL) methods offer a range of strategies for capturing useful representations from music audio without the need for labeled data. While some techniques focus on preserving comprehensive details through reconstruction, others favor semantic structure via contrastive objectives. Few works examine the interaction between these paradigms in a unified SSL framework. In this work, we propose a multi-view SSL framework for disentangling music audio representations that combines contrastive and reconstructive objectives. The architecture is designed to promote both information fidelity and structured semantics of factors in disentangled subspaces. We perform an extensive evaluation on the design choices of contrastive strategies using music audio representations in a controlled setting. We find that while reconstruction and contrastive strategies exhibit consistent trade-offs, when combined effectively, they complement each other; this enables the disentanglement of music attributes without compromising information integrity.

\end{abstract}

\section{Introduction}
Audio representation learning methods aim to learn meaningful features from input audio signals through pre-training tasks. 
Depending on the pretext task, models tend to prioritize different properties of the learned representation space, including
preserving the full complexity of the input signal, organizing embeddings semantically, or disentangling factors of variation %
within the latent space.

Reconstructive strategies such as autoencoding \cite{huang2022masked, baevski2020wav2vec20frameworkselfsupervised, liu2024semanticodec, chong2023masked, baade2022maeastmaskedautoencodingaudio} incentivize the model to compress audio signals into a compact yet informative representation to enable accurate reconstruction.  %
Contrastive learning (CL) exploits the assumption of commonalities across different views of data with the same semantics\cite{oord2018infonce, chen2020simclr, zbontar2021barlow, bardes2022vicregvarianceinvariancecovarianceregularizationselfsupervised, wang2021multiformatcontrastivelearningaudio}, such as augmented views of the same audio clip \cite{cola, fonseca2021unsupervised, al2021clar}.
It follows that such approaches often learn high-level semantics that excel at discriminative tasks by filtering out trivial variances--information that reconstructive models would preserve.
Both reconstructive and contrastive strategies produce effective representations for downstream tasks such as audio classification and retrieval.
However, combining their strengths remains an open challenge in the music audio domain.

Recent methods have shown the promising potential of integrating one strategy within another. Methods such as \cite{gong2023contrastive, jiang2020speech} jointly optimize contrastive and reconstruction losses to improve semantic correspondence. %
Others leverage both strategies to improve representation structure and achieve disentanglement. %
For example, \cite{xiao2021looc, liu2023focal, eastwood2023disaug, guinot2024leave, mccallum2024similar} 
maintain both view invariances and variances in disentangled subspaces to mitigate information loss in traditional CL methods. %
Some autoencoding methods focus on disentangling factors in independent latent variables %
by leveraging distributional assumptions %
\cite{chen2018btcvae, kim2018fvae, mathieu2019disdis, brima2024syntone}. 

In the music domain, %
many works have also explored combinations of both strategies to tackle pitch-timbre disentanglement. Some incorporate contrastive objectives directly within reconstruction frameworks \cite{luo2020unsupervised, tanaka2021pitch}, while others design augmentations for multi-stage generative pipelines\cite{tanaka2025unsupervised}, or construct paired views in a multi-view learning setting \cite{wilkins2024selfsupervisedmultiviewlearningdisentangled} without the explicit use of contrastive terms. %
These directions raise interesting questions about how different SSL strategies, namely contrastive and reconstructive, can be utilized within the same embedding space to promote informativeness, semantic structure, and disentanglement simultaneously. 
Yet in many cases, design decisions about how to apply each objective are implicit and not well-studied, something recently explored in music tagging \cite{meseguer2024experimental}.

Motivated by recent developments in unifying both paradigms, we design a multi-view learning framework to investigate how the interaction of reconstructive and contrastive objectives affects the representation space. 
The multi-view setup naturally affords the possibility to separate common and unique attributes across views, creating an embedding space composed of semantically-structured subspaces.
We use contrastive techniques to design objectives serving as the disentanglement incentive: not only invariance terms that learn commonalities like traditional CL, but also divergence terms that separate specificity. %
We systematically study this interaction between contrastive and reconstructive objectives with different design choices via downstream pitch and instrument classification tasks in a controlled setting.
Our findings show that optimizing for one desirable property (e.g., information preservation or disentanglement) often comes at the expense of another, reflecting trade-offs in self-supervised representation learning. Within the scope of our study, the evaluation highlights this tension, offering useful insights into future directions for unified SSL frameworks for audio.
Our contributions are threefold: 
\begin{itemize}
    \item We propose a multi-view SSL disentanglement framework for music audio representation learning to investigate the interactions between contrastive and reconstructive principles.  
    
    \item We design objectives for multi-view disentanglement as an extension of traditional contrastive losses and analyze behavior patterns of different design choices.
    \item We identify intriguing trade-offs between disentanglement and information preservation through a systematic evaluation
    and provide insights into a %
    design recipe that balances both.
\end{itemize}

\begin{figure*}[t]
  \centering
  \centerline{\includegraphics[width=0.88\linewidth, trim={1cm 0.5cm 1cm 1cm}]{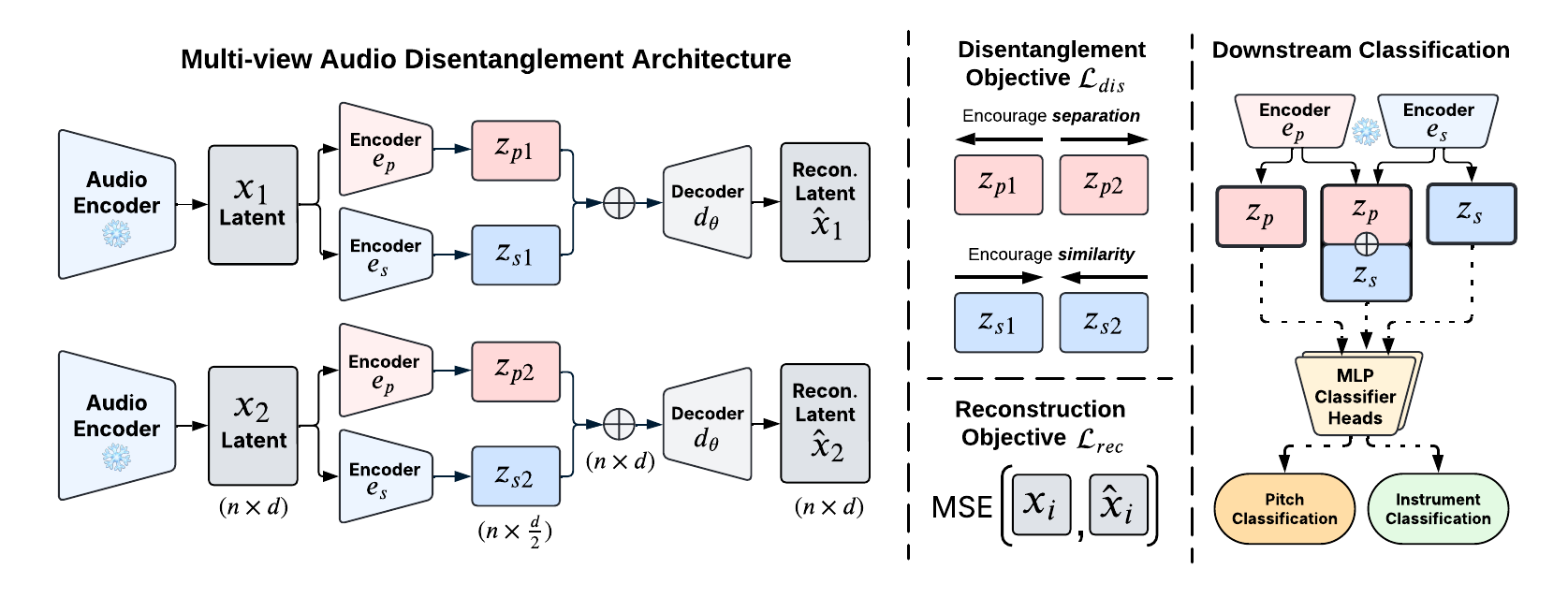}}
  \caption{Our self-supervised multi-view audio representation learning framework incorporates both disentanglement and reconstruction objectives. Learned factorized latents are used for downstream classification tasks.}
  \label{fig:block_diag}
\end{figure*}

\section{Method}
We propose a self-supervised multi-view framework that combines reconstructive and contrastive principles to investigate the disentanglement of musical attributes in the latent space. We focus on pitch and instrument latent disentanglement in this study, where we pass views of data (e.g., audio clips) with the same instrument but different pitch as input.
We aim for our model to learn features \textit{common} to both views (e.g., instrument) in the \textit{shared} latent space, and \textit{unique} features to each view (e.g., pitch) in the \textit{private} space, encouraging semantically structured representations across distinct subspaces.

\subsection{Autoencoder Framework}%
Our multi-view autoencoder framework is shown in Figure \ref{fig:block_diag} for the instance of $2$ views of data. Given an input audio spectrogram, we pass the input through a frozen pre-trained audio encoder. This yields a 2D spatiotemporal latent $x_i$ with dimension $(n \times d)$, where $i$ is the data view, $n$ is the temporal dimension, $d$ is the embedding dimension.

We use AudioMAE \cite{huang2022masked} as the pretrained audio encoder for this study. We performed UMAP \cite{mcinnes2018umap} clustering of embeddings of NSynth \cite{nsynth2017} audio clips from a variety of pre-trained encoders (AudioMAE\cite{huang2022masked}, AudioLDM \cite{liu2023audioldm}, DAC \cite{kumar2023highfidelityaudiocompressionimproved}) alongside log-mel spectrograms and found that AudioMAE showed the strongest clustering trends in terms of pitch and instrument.
We pass the AudioMAE latent $x_i$ through two encoders: $e_p$ (private) and $e_s$ (shared). 
Both $e_p$ and $e_s$ are $2$-layer MLPs with ReLU activations. The output of $(e_p(x_i))$ is the private latent $z_{pi}$ and the output of $e_s(x_i))$ is the shared latent $z_{si}$, both of dimension $(n \times \frac{d}{2})$. Note that $e_p$ and $e_s$ share weights across views.

Lastly, the private $z_{pi}$ and shared latent $z_{si}$ per view are concatenated and sent to the decoder $d_{\theta}$, which also shares weights across views. The decoder reconstructs the input AudioMAE latent, forming $\hat{x}_i$. We apply a mean-squared error (MSE) reconstruction objective, denoted $\mathcal{L}_{rec}$, to encourage the  learned latent space (e.g., $z_{pi} \oplus z_{si}$) to capture the complete information of the input.

\subsection{Disentanglement Objectives} 
\label{sec:dis_obj}
To encourage the model to learn disentangled representations, we experiment with three families of contrastive principles: InfoNCE, cosine similarity, and VICReg. Each defines a way to compute alignment or divergence between latent representations. Within each family, we experiment with two types of disentanglement objectives in the latent space: \textbf{similarity}-based objectives promote alignment between shared latents across paired views, reinforcing the encoding of invariant factors (e.g., instrument identity), and \textbf{separation}-based objectives, which incentivize divergence between private latents, encouraging the encoding of varying factors (e.g., pitch). 
Together, the shared and private subspaces should capture semantically distinct yet complementary attributes of the input. 
We use $\mathcal{L}_{dis}$ to denote the disentanglement loss, which can take a specific combination ($c^\phi$) of the contrastive principle $c$ and objective type $\phi$:, where \textit{$c \in \{\text{InfoNCE}, \text{Cosine}, \text{VICReg}\}$}, \textit{$\phi \in \{\text{SIM}, \text{SEP}, \text{SIM+SEP}\}$}:
\begin{itemize}
    \item \textbf{$c^{\text{SIM}}$}: similarity objective applied to shared latents $z_{s1}$ and $z_{s2}$.
    
    \item \textbf{$c^{\text{SEP}}$}: separation objective applied to private latents $z_{p1}$ and $z_{p2}$.
    \item \textbf{$c^{\text{SIM+SEP}}$}: linear combination of $c^{\text{SEP}}$ and $c^{\text{SIM}}$.
\end{itemize}

\textbf{InfoNCE}: 
Following traditional CL \cite{oord2018infonce},
we treat the shared latents of naturally occurring paired samples %
as positives and those of
all non-matching samples
in the batch as negatives; this is denoted  $\text{InfoNCE}^{\text{SIM}}$. 
We extend this concept to private latents as $\text{InfoNCE}^{\text{SEP}}$, using %
all private latents as negatives within a batch.
    The loss functions for both methods are defined as:
    \begin{equation}
        \small
        \mathcal{L}_{\text{InfoNCE}^{\text{SIM}}}=-\mathbb{E}_{j\in[1,B]}\left[\log \frac{\exp(\textit{sim}(e_s(x_1^j), e_s(x_2^j))}{\sum_{k}^B \exp (\textit{sim}e_s(x_1^j), e_s(x_2^k))}\right]
    \end{equation}
    \begin{equation}
    \mathcal{L}_{\text{InfoNCE}^{\text{SEP}}}=-\mathbb{E}_{j,k\in[1,B]}[\log (1-\textit{sim}(e_p(x_1^j), e_p(x_2^k)))]
    \end{equation}
    
    where $x_1^j$ and $x_2^k$ are the $j$-th and $k$-th samples from the first view and second views, respectively, in a batch of size $B$. Within each batch, two views with the same sample index $(x_1^j, x_2^j)$ form a positive pair and all other combinations $(x_1^j, x_2^k), j\neq k$ are the negatives.
    
\textbf{Cosine Similarity}: %
We apply a cosine similarity objective derived from the non-contrastive objective family (e.g., BarlowTwins\cite{zbontar2021barlow}, BYOL \cite{grill2020bootstrap}) where we do not use %
comparisons within batch. We first normalize $z_p$ and $z_s$ across batch to unit vectors. In $\text{Cosine}^{\text{SIM}}$, we maximize the cosine similarity between shared latents towards $1$. For $\text{Cosine}^{\text{SEP}}$, we minimize the cosine similarity between private latents towards $0$.  The loss functions are defined as follows:
    
    \begin{equation}
    \small
    \mathcal{L}_{\text{Cosine}^{\text{SIM}}}=-\mathbb{E}_{j\in[1,B]}\big[\log \textit{sim}(e_s(x_1^j), e_s(x_2^j))\big] 
    \end{equation}
    \begin{equation}
    \mathcal{L}_{\text{Cosine}^{\text{SEP}}}=-\mathbb{E}_{j\in[1,B]}\big[\log (1-\textit{sim}(e_p(x_1^j), e_p(x_2^j)))\big]
    \end{equation}
    
\textbf{VICReg}: %
We use VICReg \cite{bardes2022vicregvarianceinvariancecovarianceregularizationselfsupervised} in its original formulation for the shared subspace as our $\text{VICReg}^{\text{SIM}}$ method to optimize a triple objective-- \textit{invariance}: minimizing mean-squared error between paired representations; \textit{variance}:  encouraging diversity of latent variables across batch; \textit{covariance}: preventing information collapse by reducing dimension-wise redundancy within latent. %
For the private subspace, we construct $\text{VICReg}^{\text{SEP}}$ by retaining the \textit{variance} and \textit{covariance} terms but negating the \textit{invariance} term for separation.

\section{Experimental Design}
\subsection{Dataset \& Data Pairing}
We use the NSynth dataset \cite{nsynth2017} due to its well-defined generative factors. Each 4-second audio clip contains a single musical note with a unique pitch, instrument, and other factors. Our method requires a single condition of commonality between views of data, and for this study we use \textit{instrument} as that common factor. We create pairs of audio samples from the same instrument family, assuming differences in all other factors. 
Note that while we use a common factor label for data pairing, in training and at inference time, our method is fully self-supervised and labels are not needed.

Increasing the training dataset size provided only marginal accuracy gains and no change in disentanglement trends, so we opted for a smaller subset to support more extensive ablation studies. 
We use $10k$ NSynth samples for training and $5k$ for validation, stratified by instrument family and pitch. For downstream tasks, we further partition the NSynth test split into train, validation, and test subsets. We also stratify on both pitch and instrument for these subsets, resulting in $88$ pitch classes and $10$ instrument classes for evaluation.

\begin{table}[ht]
\caption{Downstream classification: disentanglement margin ($\Delta$) and overall accuracy. Multi-view (\footnotesize{$\mathcal{L}_{rec}+\mathcal{L}_{dis}$}) uses $\text{Cosine}^{\text{SEP}}$, \footnotesize{$\lambda=0.6$}.}
\centering
\scriptsize
\begin{tabular}{l|c:c|c:c}
  & \multicolumn{2}{c|}{\textbf{Pitch Clf. Acc. $\uparrow$}} & \multicolumn{2}{c}{\textbf{Inst. Clf. Acc. $\uparrow$}} \\
 \textbf{Method} & $\Delta$Pitch & Overall  & $\Delta$Inst. & Overall \\
\midrule
\textbf{Multi-view ($\mathcal{L}_{rec} + \mathcal{L}_{dis}$)} & \textbf{0.10} & \textbf{0.71} &  \textbf{0.17} & 0.94 \\
Multi-view ($\mathcal{L}_{rec}$)  & 0.00 & 0.70 &  0.01 & \textbf{0.95} \\
\midrule
CL - Pitch (InfoNCE \cite{oord2018infonce}) & - & 0.65  & - & (0.35)  \\
CL - Instrument (InfoNCE \cite{oord2018infonce}) & - & (0.14) &  - & 0.81  \\
\midrule
AudioMAE\cite{huang2022masked}  & - & 0.67 &  - & 0.91 \\
\end{tabular}
\label{tab:main_table}
\end{table}

\subsection{Loss Weighting Ablation}\label{sec:ablation_design}
For loss balancing, to avoid a disparity in scale between $\mathcal{L}_{rec}$ and %
$\mathcal{L}_{dis}$, we introduce a configuration-specific scaling factor $\gamma$ to normalize the reconstruction loss magnitude. This allows us to more fairly explore the trade-off between reconstruction and disentanglement. We sweep $\lambda \in \{0.0,\ 0.2,\ 0.4,\ 0.6,\ 0.8,\ 1.0\}$, where $\lambda$ is the weight of the reconstruction loss term and $(1-\lambda)$ is the disentanglement weight:
\begin{equation}
    \mathcal{L}_{total} = \gamma \cdot \lambda \cdot \mathcal{L}_{rec} + (1 - \lambda) \cdot \mathcal{L}_{dis}
\end{equation}

Here, $\lambda=0$ indicates using only the disentanglement loss (no decoding) and $\lambda = 1$ uses reconstruction only. We train one model per combination of $\lambda$, contrastive principle $c$, and disentanglement objective type $\phi$,
leading to 54 total experiments, where a single configuration is, for example, $\text{VICReg}^{\text{SEP}}$ for $\lambda=0.2$.

\subsection{Training Parameters}
Each model is trained for 100 epochs with batch size 16, learning rate $1e-4$, and AdamW optimization with $1e-3$ weight decay on a V100 GPU. %
We select the best models to use for downstream classification based on total validation loss.

\subsection{Downstream Classification} 
We use downstream classification to assess the quality and disentanglement of the learned latent spaces. %
As shown in the right panel of Fig.~\ref{fig:block_diag}, after training the multi-view disentanglement model, we freeze the private and shared encoders, $e_p$ and $e_s$, and use them as feature extractors to obtain latents for two downstream tasks: pitch classification and instrument family classification. We train independent classification heads on top of the private ($z_p$), shared ($z_s$), and concatenated latents ($z_{p} \oplus z_{s}$) for each downstream task. Each classification head consists of a single linear layer with $50\%$ dropout, trained with Cross Entropy Loss.

\subsection{Evaluation Metrics}
We use the following metrics to assess model performance in terms of disentanglement, learned semantics, and overall information retention:

\textbf{Subspace-level accuracy per-task}: accuracy when using either $z_p$ or $z_s$ individually as the input feature for downstream classification on pitch and instrument classification, capturing per-subspace semantics. %

\textbf{Overall accuracy per-task}: downstream classification accuracy when using $z_p \oplus z_s$. This measures the overall informativeness of the learned embeddings, regardless of disentanglement. This uses a latent of double the size of the subspace-level metrics due to concatenation. 

\textbf{Disentanglement margin per-task}: if there is no disentanglement in the latent space, we expect classification performance to be the same when using either $z_p$ or $z_s$ as downstream features. Thus, we use the \textit{difference} ($\Delta$) between subspace-level accuracies as a proxy for disentanglement, targeting a positive margin for each task:
    \begin{itemize}
         \item \textbf{$\Delta \text{Pitch}$} $= \text{PitchClf}(z_p) - \text{PitchClf}(z_s)$
        \item \textbf{$\Delta \text{Instrument}$}  $= \text{InstrumentClf}(z_s) - \text{InstrumentClf}(z_p)$
    \end{itemize}

\textbf{Mean-squared error (MSE)}: We compute the MSE between the original and reconstructed latents, where a lower MSE indicates a more accurate reconstruction. MSE combined with overall accuracy give us a holistic view of overall information completeness.%

\begin{figure}[t]
  \centering
  \centerline{\includegraphics[width=1\columnwidth, trim={0 0 1cm 2cm}]{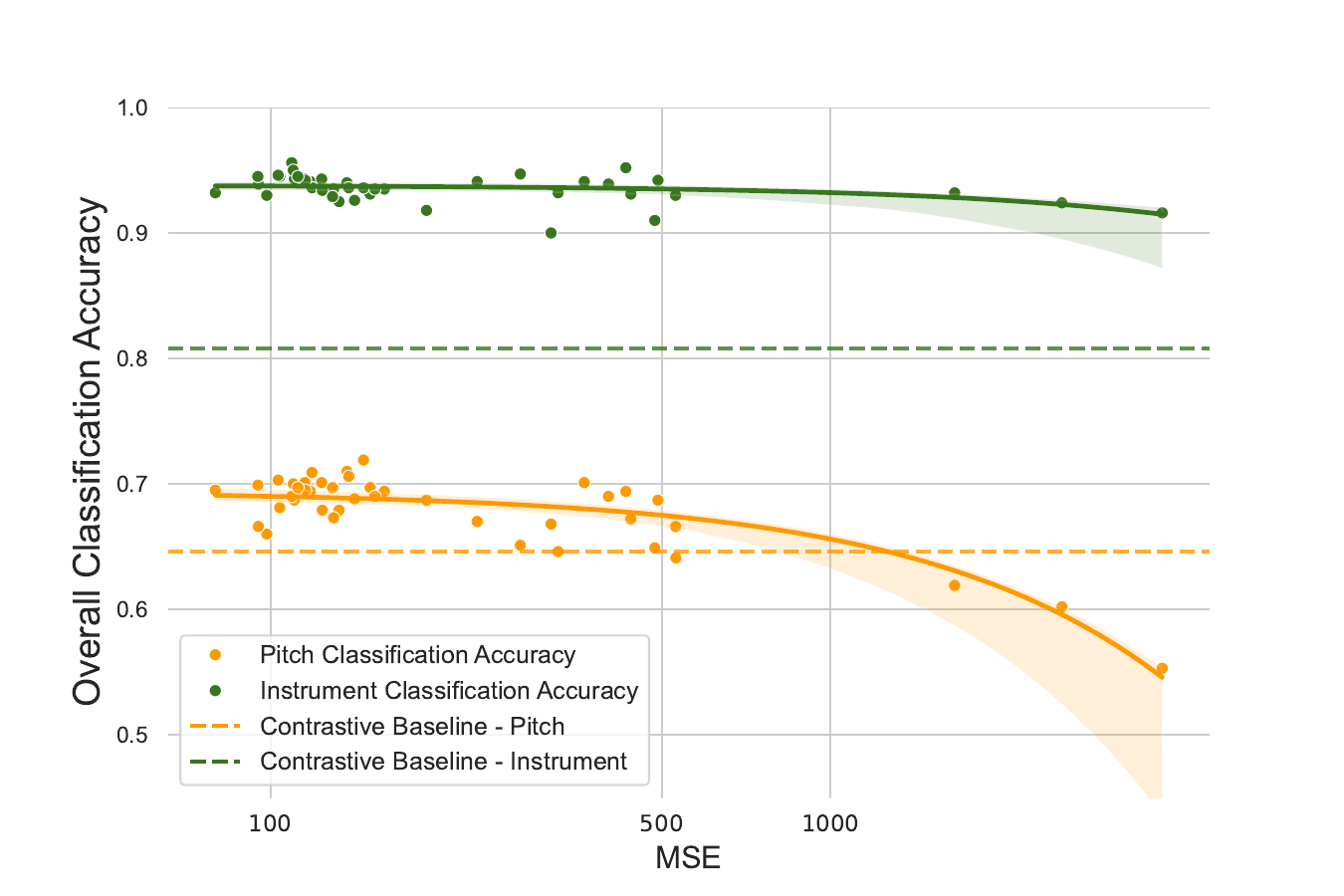}}
  \caption{Overall downstream classification accuracy vs. reconstruction MSE for all model configurations using concatenated latents. 
  }
  \label{fig:MSE_acc}
\end{figure}

\subsection{Baseline Methods}
For our baseline methods, we compare with pared-down versions of our proposed approach and with classical CL. Each method uses AudioMAE frozen latents as input.
\textbf{AudioMAE (frozen)}: Uses off-the-shelf AudioMAE embeddings directly in downstream training. %
\textbf{Reconstruction Only}: Our multi-stream autoencoder architecture, trained solely with $\mathcal{L}_{rec}$. \textbf{Contrastive Learning}: Standard CL approach using only the shared encoder from our architecture (Fig. \ref{fig:block_diag}), producing $z_{s1}$ and $z_{s2}$. We apply an InfoNCE loss to align pairs of representations controlled by either pitch or instrument as a common factor. No decoder is used. The size of $z_{s1}$ and $z_{s2}$ is doubled to create a fair comparison with overall accuracy from our system.

\section{Results and Discussion}

Our framework allows us to analyze the tension between disentangled representations and preserving complete input information, revealing that disentanglement in audio latent spaces often entails trade-offs.

\subsection{Main Comparison}
Our optimal model configuration in terms of overall information fidelity and disentanglement 
uses the $\text{Cosine}^{\text{SEP}}$ disentanglement objective jointly with the reconstruction objective, using $\lambda = 0.6$. In Table \ref{tab:main_table} we show downstream results of this system relative to baseline methods. Our multi-view approach creates semantically disentangled subspaces-- shown by our positive disentanglement margins, \textit{without} sacrificing overall accuracy of the full latent representation-- shown by comparison with the system using only the reconstruction objective. Importantly, we demonstrate that our method outperforms the CL baselines in terms of overall accuracy per-task, improving upon the CL-pitch baseline by 6pp and CL-instrument by 13pp. Simultaneously, we boost the disentanglement margin of the $\mathcal{L}_\text{rec}$-only baseline by 10pp in $\Delta$Pitch and 16pp in $\Delta$Instrument with our hybrid approach.

Using AudioMAE embeddings directly in downstream training yields a strong baseline for downstream task accuracy, though in a fully entangled form. %
We find that applying CL to the AudioMAE space directly hurts performance in comparison to AudioMAE off-the-shelf, possibly due to batch size or architecture limitations.

\begin{figure}[t]
  \centering
  \centerline{\includegraphics[width=1.1\columnwidth, trim={0 0 0 2cm}]{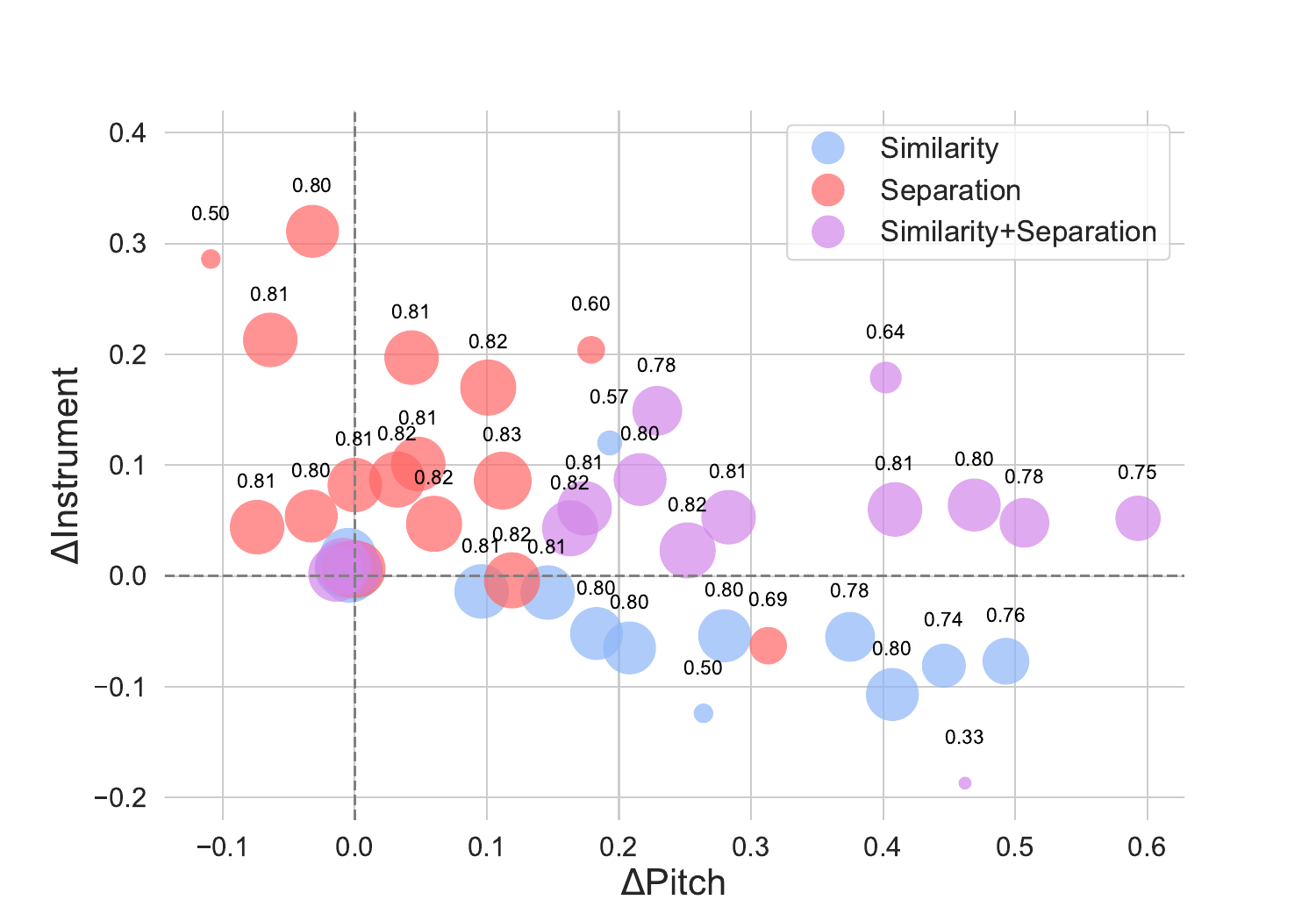}}
  \caption{%
  Disentanglement margins colored by objective types (similarity, separation, and both) measured per model configuration.
  Size reflects average downstream overall accuracy across tasks.} 
  \label{fig:bubbles}
\end{figure}

\subsection{Reconstruction Boosts Downstream Accuracy}
We observe in Figure \ref{fig:MSE_acc} that, %
regardless of model configuration, 
better reconstruction, indicated by lower MSE, is generally correlated with higher overall downstream classification accuracy. We also observe that our method, trained with both reconstruction and disentanglement incentives, consistently outperforms the CL baseline across configurations. The addition of a reconstruction objective improves our model's capacity to capture richer signal information in the latent space compared to CL, affording the ability to disentangle the latent space without information loss. Data points in Fig. \ref{fig:MSE_acc} with higher MSE and lower accuracy are primarily model configurations with a lower weight $\lambda$ applied to $L_{rec}$.

\subsection{Separation is Key for Disentanglement}

In Figure \ref{fig:bubbles}, we analyze the trade-off between disentanglement quality and classification accuracy when using similarity (blue) or separation-based (red) objectives, or a combination of both (purple). Each point is a model configuration of contrastive principle $c$, disentanglement objective type $\phi$, and reconstruction weight $\lambda$. Size represents average overall classification accuracy across tasks.
Ideal configurations are larger circles in the upper-right quadrant, indicating positive disentanglement margins and higher accuracy across tasks. 

We found that the choice of contrastive principles did not yield consistent disentanglement trends; rather, the choice of a similarity or separation-based objective proved much more influential. Given that the $\mathcal{L}_{rec}$-only baseline has an average accuracy of $0.82$ and disentanglement margins of $0$, in Fig. \ref{fig:bubbles} we show that for multiple model configurations, we are able to achieve some disentanglement in the latent space, without losing information per this baseline. 

Crucially, we observe that separation-based objectives are more effective than similarity in terms of achieving positive disentanglement for both tasks and comparable overall accuracy to the $\mathcal{L}_\text{rec}$-only baseline. Similarity-based approaches 
struggle to produce positive disentanglement margins for instrument classification and sacrifice overall accuracy versus the $\mathcal{L}_\text{rec}$-only baseline.
Further, the models using both mechanisms almost universally achieve positive disentanglement margins on both tasks while maintaining competitive accuracy levels. Lastly, we observe that when disentanglement margins are highest along either axis, overall information preservation is compromised.

\begin{figure}
    \centering
    \includegraphics[width=\columnwidth, trim={0 0 0cm 3cm}]{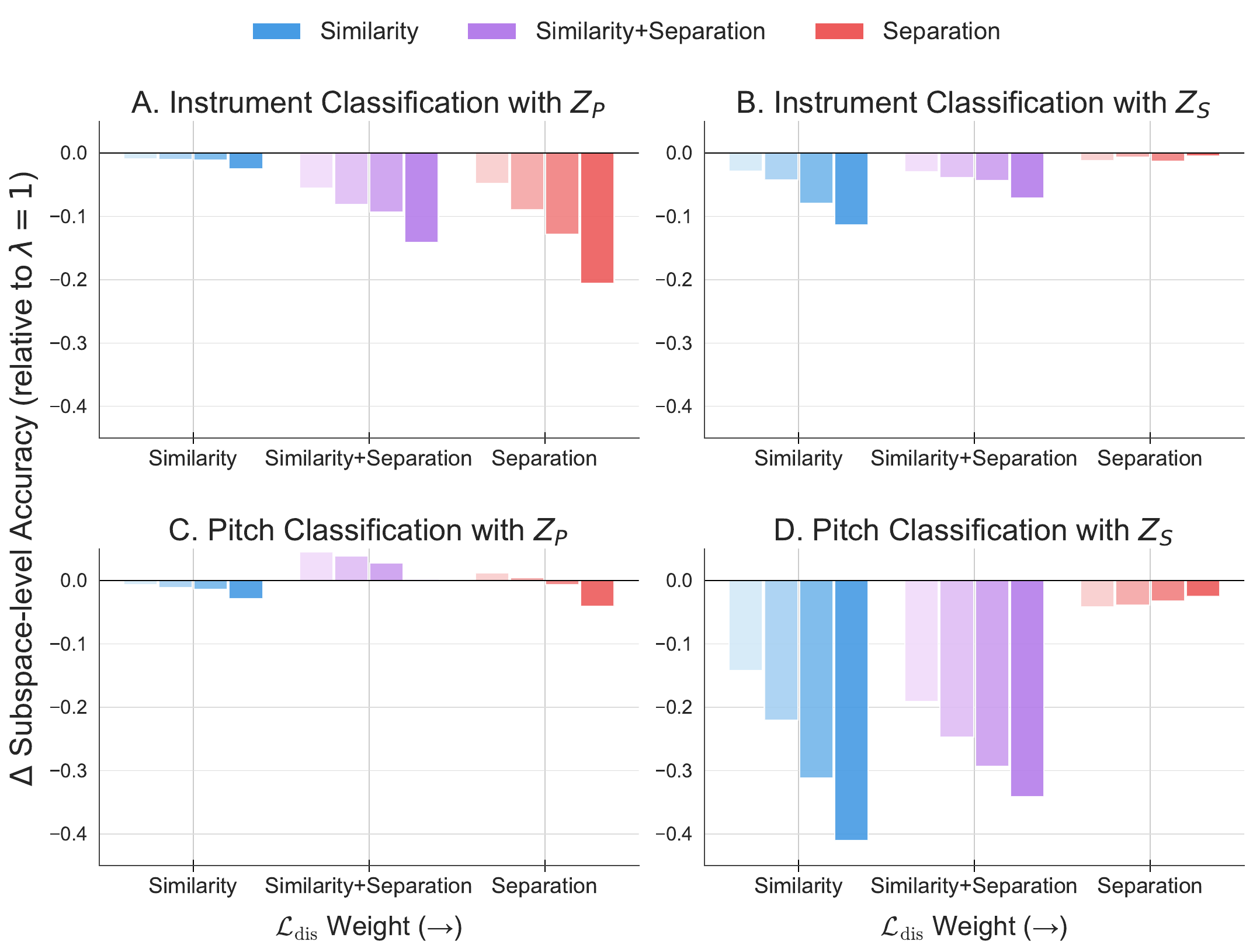}
    \caption{$\Delta$Subspace-level accuracy using $z_p$ or $z_s$ as downstream features, relative to reconstruction-only models ($\lambda=1)$. Bars are averaged across contrastive principles. The color gradient from light to dark represents an increasing weight of $\mathcal{L}_{dis}$.}
    \label{fig:quad}
\end{figure}

\subsection{Studying Subspace Information Flow}
Figure~\ref{fig:quad} illustrates how the choice of disentanglement objective type affects subspace-level accuracy. 
Across all methods, a dominant trend of accuracy decrease from the $\mathcal{L}_\text{rec}$-only baseline emerges, because shifting the objective priority from reconstruction to disentanglement always leads to information loss. However, the extent of such a trend varies depending on objective types and downstream configurations.

Comparing subplots (\textit{A}, \textit{C}) and (\textit{B}, \textit{D}) in Fig. \ref{fig:quad}, we first observe that when applying $\mathcal{L}_{dis}$ to a given subspace, performance is relatively unaffected in the \textit{other} subspace. This confirms that only the optimized subspace shows an effect when used downstream.

Next, we find a surprising trend consistent across weights of $\mathcal{L}_{dis}$: using separation-based objectives (on $z_p$) has the largest effect on \textit{instrument} classification with $z_p$ (\textit{A} in Fig. \ref{fig:quad}), and conversely similarity-based objectives on $z_s$ have the biggest impact on \textit{pitch} classification with $z_s$ (\textit{D} in Fig. \ref{fig:quad}). This at first seems counterintuitive; e.g., one might expect that encouraging the shared latents to contain similar information would show the largest impact in instrument classification using $z_s$, given that the \textit{instrument} information is shared across views by design. We hypothesize that this trend actually explains the disentanglement behavior; each objective succeeds in squeezing irrelevant information out of the optimized subspace.

\section{Conclusion}
We propose a multi-view SSL framework for disentangled music audio representation learning, designed to explore how contrastive and reconstructive strategies interact. 
We highlight the tension between information preservation and disentanglement, finding that while one often comes at the expense of the other, this balance is achievable with an informed objective design.
Extensive experiments on the NSynth dataset demonstrate that while separation-driven objectives are crucial for effective disentanglement, combining them with similarity-based objectives yields a consistent balance between factor disentanglement and information preservation.

Future work will explore the generalizability of such findings to less-structured audio domains, such as cover song datasets where distinct factors may be more abstract or labels are absent. This aims to further validate the robustness and flexibility of disentangled SSL frameworks in more diverse and realistic music settings.

\section{Acknowledgments}
\label{sec:ack}
We thank our colleagues at Bosch for their support on this work, which is partially supported by award $\#$24-2283 (Robert Bosch, LLC).

\clearpage
\bibliographystyle{IEEEtran}
\bibliography{refs25}

\end{document}